\documentclass{article}
\usepackage[a4paper, total={7in, 9.5in}]{geometry}
\usepackage{amsmath}
\usepackage{graphicx}
\usepackage{hyperref}
\usepackage[english]{babel}
\usepackage{natbib}
\usepackage{xcolor-material}
\usepackage{ulem}
\usepackage[T1]{fontenc}
\usepackage{ae,aecompl}
\usepackage{newtxtext,newtxmath}
\usepackage{soul,xcolor}
\setstcolor{red}

\begin{document}
\begin{flushleft}
\Large{\bf{The Impact of Solar Radiation on the Martian Upper Atmosphere}} \\
\vspace{5mm} 
\large{S.\,C. Chakravarty$^{1,2}$ and Kamsali Nagaraja$^{2}$} 
\\
\small{$^{1}$Indian Centre for Space Physics (ICSP), Kolkata 700099, Email:subhas@csp.res.in or chakravartysubhas@gmail.com (Chakravarty)\\
$^{2}$Department of Physics, Bangalore University, Bengaluru 560056, Email:kamsalinagaraj@bub.ernet.in (Nagaraja)}\\ 
\end{flushleft}

\large{\bf\noindent{Abstract}}\\

\large{The first in-situ measurements of the altitude profile of Martian upper atmospheric density and composition were carried out by the Viking lander missions in 1976. The MAVEN and MOM spacecraft launched in September 2014 with mass spectrometers and solar radiation measuring payloads have vastly expanded this initial data base. Using a rare set of near-simultaneous data from these two orbiters, we find that there is either an increasing (e.g., for $CO_2$ and $Ar$) or a decreasing (e.g., for $O$) trend of the density profiles by a factor of 2 between June 1 to June 15, 2018 in the height region of $\sim$150-300 km. A time series analysis of the concurrent in-situ solar EUV spectral flux and the $H^+$ ion velocities of the incident solar wind measured near MAVEN periapsis showed the former going through a decrease of only $\sim$10\% compared to the latter's decrease by a factor of 4 within the same non-solar-flare period of observation. The estimates of standard errors and the use of the linear regression analysis for the correlation coefficients between densities and solar radiation components have been carried out. Invoking simple photochemical equilibrium conditions with the dissociation of $CO_2$ (producing $CO$ and $O$) through solar EUV radiation and the solar wind $H^+$ ion impact process, the day-to-day variations of these constituents are estimated. The high and significant anti-correlation between the density variations of $CO_2$ and $O$ due to the dissociation of $CO_2$ by the solar wind particle radiation is clearly demonstrated. The cause for the increasing densities of $Ar$ like that of $CO_2$ during this period is more complex and would likely be governed by the temperature variations due to absorption of solar EUV/charged particle radiation and other interacting dynamical effects.\\

\textnormal

\section{Introduction}

Considerable progress has been made to conduct in-situ observations of Martian surface and atmospheric parameters using orbiters, landers and rovers. Near-surface meteorological data has been analysed and consolidated in diurnal, seasonal and inter-annual variations [1]. However, till recently, the measurements of upper atmospheric composition and density of Mars have been limited to the two sets of observations taken by the Viking landers while traversing down through its thin atmosphere [2-4]. In September 2014, the Mars Atmosphere and Volatile Evolution (MAVEN) and the Mars Orbiter Mission (MOM) spacecraft entered Martian orbit and were placed in elliptical orbits around Mars with one of the main objectives to gather substantial data on spatial and temporal profiles of various upper atmospheric neutral/ion density and composition parameters [5,6].

Mars has a well-mixed region of the homosphere with the homopause at $\sim$120\,km altitude. The thermosphere extends above 120\,km finally merging into the exosphere. From the exosphere the lighter gases may escape due to negligible neutral gaseous collisions and through interaction with solar EUV and energetic charged particle radiations. This loss process usually starts from around 220 km, called the exobase [7,8]. With the availability of continuous thermosphere/exosphere density data (due to its lower perigee) for more than one Martian year from MAVEN, it is possible to study the effect of solar forcing on neutral densities [9]. 

The results obtained from both MAVEN using NGIMS (Neutral Gas and Ion Mass Spectrometer) payload and MOM using MENCA (Mars Exospheric Neutral Composition Analyser) payload have so far provided basic information about the spatial variation of the upper atmospheric gas constituents and ion species delineating their vertical and horizontal distributions [10-12]. Recently Sarris [13] has suggested that short term neutral density changes in Earth's thermosphere could be attributed to the variations of solar EUV as well as to the solar particle radiation. In the absence of an earth-like magnetosphere, the charge particle interaction with Martian atmospheric constituents would need further studies.

MAVEN has many instruments which measure solar wind parameters along its track covering the lower altitude range around the periapsis ($\sim$150-300\,km) of MAVEN, which is the main height region of interest in this paper. In one day, this region near perigee is covered $\sim$5 times and the average flux received each day depends on the daily variation of solar activity.
The primary purpose of this paper is to explore the non solar-flare variation of different thermosphere-exosphere gas constituents of Mars by using NGIMS/MAVEN and MENCA/MOM data and to assess the role of the variable solar energetic radiation as a possible cause.

\section{Dataset and Method of Analysis}

In this study, we mainly consider the region of the Martian atmosphere between the upper thermosphere and lower exosphere ($\sim$150-300 km) with the exobase level at $\sim$220 km. This region is identified as the space where the interaction of solar EUV and charged particle radiation with various atmospheric constituents takes place. Enhanced solar radiation levels may lead to the escape of $H$ and $O$, due mainly to the relatively weak surface gravity of Mars as compared to that of the Earth [14].

While both spacecraft measured the upper atmospheric composition and densities, their respective spatial and temporal coverage are quite different with reference to the altitudes of interest. We could only select the period  of June 2018 with near simultaneous observation in the height range of interest. Such a coincidence of getting near-simultaneous observations is very rare and has happened for the first time.

Further, the period of only June 1-15, 2018 has been selected and the second half of June 2018 avoided, which was affected by the Planet Encircling Dust Event (PEDE). The effect of the global dust storm on thermospheric densities has been studied using the available MENCA and NGIMS data for the event during the second half of June 2018 which demonstrated the asymmetry between the daytime and nighttime thermospheric density observations of both the spacecraft [15].
In an earlier study we have already shown the highly sensitive response of the neutral atmospheric composition and density to an eruptive event of coronal mass ejection (CME) using MENCA data for December 2015, when MAVEN data was not available [12].

The results in this paper are based on the solar quiet time vertical atmospheric density profiles of constituents, $CO_2$, $O$, $Ar$, $N_2$ etc., derived from similar mass spectrometric instruments carried by both the spacecraft. As the time interval between two successive observations is a few days for MOM and only a few hours for MAVEN, we have mainly used MAVEN data for better statistics.    

\subsection{MENCA Instrument}
MOM arrived at Mars on 24 September 2014 in an eccentric orbit of $\sim$422\,km$\times$76,993\,km with an orbital period of about 72\,h. During December 2014, orbital manoeuvres brought down the periapsis altitude to around 263\,km. The MENCA observations measure total atmospheric pressure and partial pressures of various atmospheric constituents covering 1-100 amu.  More details about the instrumentation, limitations, observation errors and sources of contamination can be found in Bhardwaj et al. [16-18].

The data from this experiment has been made available for the project from time to time through the ISRO Space Science Data Centre (ISSDC). It consists of total pressure and partial pressure values as counts in ampere units with a variable time resolution of 12-48\,s near periapsis. The inbound and outbound trajectories cover the lowest altitudes. Before this base-level data can be used for scientific studies, further processing has been done by us, which include instrument calibration to convert the raw current counts to pressure unit Torr, background correction with respect to higher altitude measurements, temporal/spatial smoothing and tagging each observation point with ancillary data such as latitude, longitude, altitude and solar zenith angle.  

The data processing for each orbit involves handling of a number of file pairs, each containing the total atmospheric pressure and partial pressures of constituents for different time blocks. The files for partial pressures contain the data in a continuous time-sequenced array form, which are converted into a tabulated columnar format with a time of observation in UTC, corresponding to the nine values of partial pressures for each constituent from 1 to 100 amu and total pressures synchronised in time. ISSDC also provides the specific SPICE (Spacecraft Planet Instrument C-matrix Events) kernel files to extract the local solar time (LST), altitude, latitude, longitude and solar zenith angle values linked to each record of observation. The calibrated partial pressure values in Torr along with the associated ephemeral and spatial information for each orbit have been used in  analysing the results. After the above treatment of basic data, the MOM orbit-wise analysis is carried out to select the useful partial pressure values from the periapsis altitude to about 500\,km.

\subsection{NGIMS and other Instruments onboard MAVEN}
The NGIMS instrument of the MAVEN spacecraft has been utilised to determine the density and composition of the upper atmosphere’s neutral and ionic species in a range of 2 to 150\,amu [19]. NGIMS science-mode observation is conducted between 500\,km to the periapsis altitude of $\sim$150\,km during each orbit lasting 1200\,s with a vertical resolution of $\approx$2\,km [20]. The level-2 (version-8 and revision-1) datasets of NGIMS, retrieved from MAVEN Science Data Center are used for this study. The Extreme Ultra Violet Monitor (EUVM) instrument on MAVEN measures the solar irradiance at Mars using three photometers sensitive to the wavelengths 0.1-7\,nm, 17-22\,nm and 121.6\,nm [21]. 

The Solar Energetic Particle (SEP) instrument on MAVEN consists of two dual, double-ended solid-state telescopes with four look directions per species, optimised for parallel and perpendicular Parker Spiral viewing of energetic ions (25\,keV to 12\,MeV) and electrons (25\,keV to 1\,MeV) with 1\,s time resolution. SEP can measure energy fluxes that range from 10 to 10$^{6}$\,eV/(cm$^{2}$\,s\,sr\,eV) [22].  

The Solar Wind Ion Analyzer (SWIA) measures the energetic ions from the upstream solar wind and magnetosheath around Mars, within the energy range of 5-25000 eV\,q$^{-1}$ and an angular range of 360$\times$90 $\deg$ [23]. During its nominal operation, pointing to the Sun from the top corner deck of the MAVEN spacecraft, the SWIA instrument measures the nominal solar wind flows. This instrument provides high-resolution ion velocity measurements with a broad energy spectrum. We utilize the hourly mean velocity data from the SWIA instrument for the period June 3-15, 2018 representing the average condition of these velocity spectra through the altitudes of interest.

\section{Results and Discussion}

The daytime mean density profiles of $Ar$ measured by both the spacecraft on a few selected days during June 2018 are shown in figure-1. The profile shapes and a broadly increasing trend of density values between June 1-15, 2018 measured by both NGIMS and MENCA compare quite well as can be seen between the 160-220\,km height range within the limitation of different sampling rates and height resolutions. In the absence of any solar energetic event or any effect of dust storm sweeping the lower atmosphere during this time, such a systematic and large variation within a short period of 2 weeks is difficult to explain. For both sets of data the changes in Solar Zenith Angles (SZA) are only of the order of a few degrees during the observation period covering a small arc of the orbital path near the periapses. This is illustrated through figure-2 and figure-3 for MOM and MAVEN trajectories respectively. Hence an average density variation up to a maximum of 10\% can only be explained due to the SZA effect. The observed increase of profile densities by a factor of 2-3 (derived from figure-1b) would mean a thermospheric temperature change from 250\,K to 500\,K (estimated using the relation H = kT/mg, where H is the scale height, k the Boltzmann constant, g the gravitational acceleration of Mars and m the molar mass of $Ar$, and the hydrostatic equilibrium relation ${N = N_o \exp\left(-\frac{(z-z_o)}{H}\right)}$; where N${_o}$ and N are number densities of $Ar$ at heights z${_o}$ and z respectively). Many authors have computed the mean values of temperatures from scale heights in order to explain the density variations [18, 20, 24]. Bharadwaj et al. [18] measured $Ar$ density using MOM data and attributed sharp bends in scale height profiles due to thermospheric temperature changes which can explain the variations in $Ar$ density. In our results, the absence of any sharp change in the slope of the $Ar$ density profiles (figure-1) does not indicate a major temperature variation of the required magnitude. The small fluctuations within the individual vertical profiles point to other possible dynamical effects, such as the tidal oscillations, planetary and gravity waves propagation [25].

\begin{figure}
	{\includegraphics[width=\columnwidth]{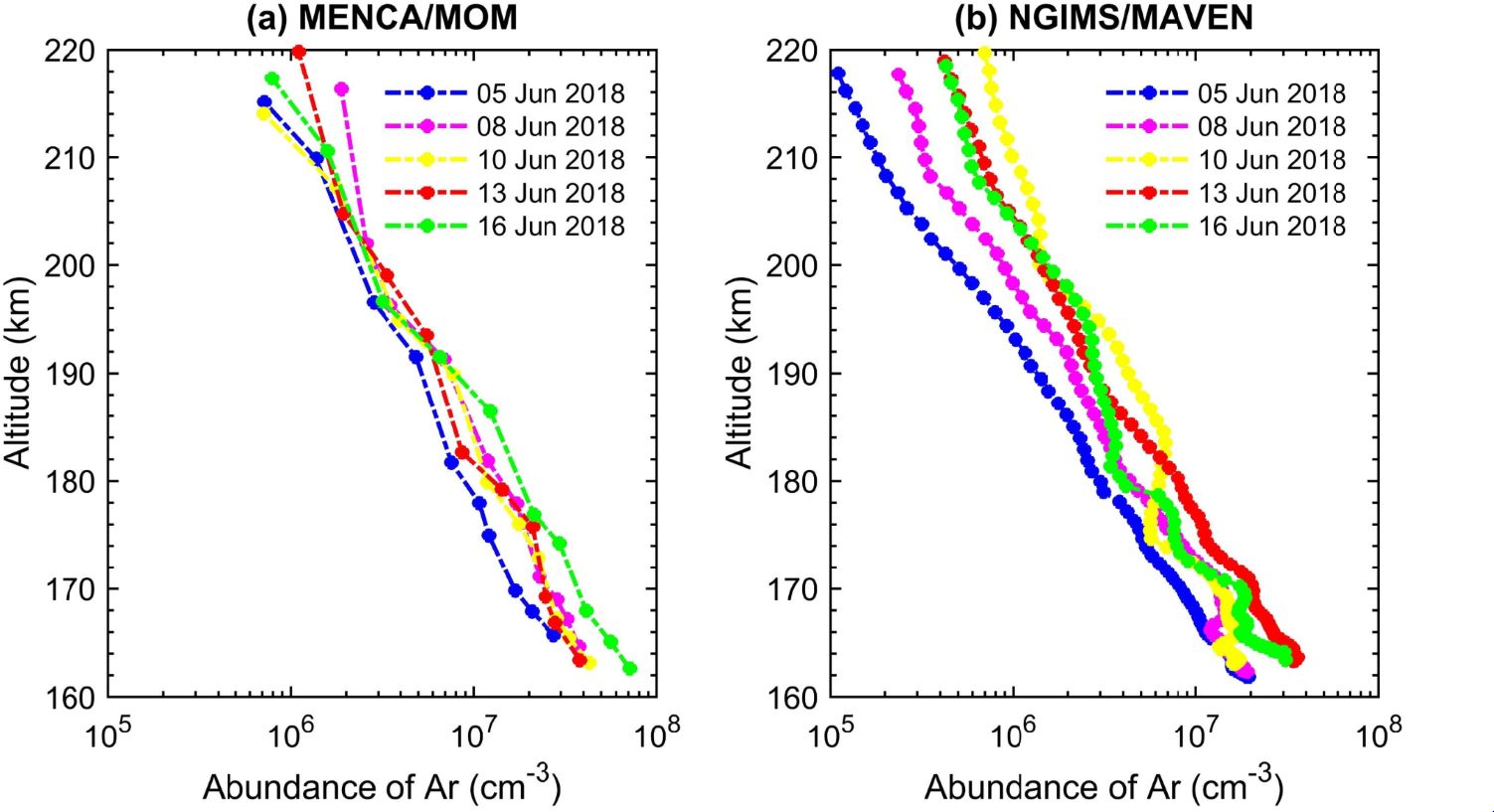}} 
	\caption{Mean daytime density profiles of $Ar$ in the 160-220 km of the upper atmosphere of Mars as measured by a) MENCA and b) NGIMS for a few selected days between June 1-15, 2018. The MENCA data consists of  processed mean values of partial densities of gas constituents with a height resolution of 5 km and the NGIMS data is obtained from the MAVEN Science Data Centre. Here the daily mean $Ar$ concentrations are computed from 5 daytime passes with a height resolution of 2\,km.}
	\label{fig001}
\end{figure}

\begin{figure}
\begin{center}
	{\includegraphics[width=0.75 \textwidth]{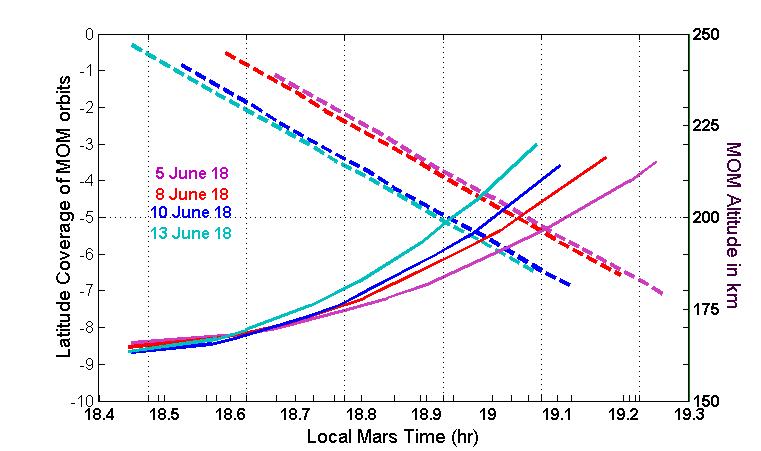}}
	\caption{LST-Latitude (dashed lines) and LST-Altitude (solid lines) plots for selected days of MOM observations during June 1-15, 2018. The density profiles of various atmospheric constituents for these four days were obtained under very similar local time and latitude conditions and hence would not be the cause for day-to-day variations. Similarly, it is found that the effect of changes in longitude and SZA would not be significant to affect any considerable day-to-day variation within an hour of LST.}
	\label{fig002} 
\end{center}
\end{figure}

\begin{figure}
\begin{center}
	{\includegraphics[width=0.75 \textwidth]{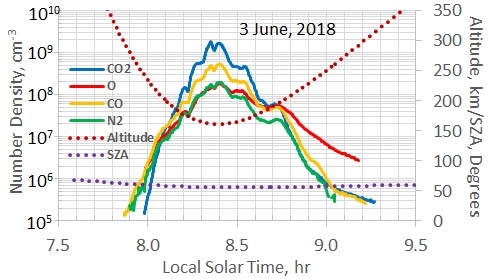}}
	\caption{Density profiles of gas constituents in the thermosphere-exosphere of Mars as measured by NGIMS (Orbit \#7154) during the local morning hours of 03 June 2018 for either side of passage through the periapsis altitude. The altitude and SZA variations are also shown as dotted lines. The height, LST and SZA all vary in a nearly similar manner for other days of observation also.}
	\label{fig003}
\end{center}
\end{figure}

Along with the SZA, LST  and latitudinal variation, figure-3 also shows the typical profiles of some of the major atmospheric constituents from one of the daytime passes of MAVEN on 3 June 2018. The plots show the measured densities along a small trace of the orbit near its periapsis ($\sim$160-300\,km). The measured values are shown for both the inbound and outbound parts of the orbital trace. For comparison with different days, we have used the densities obtained from the inbound part for further analysis. During this coverage, the LST changes only by about 40\,min and SZA by a few degrees. Plots of densities of $CO_2$, $O$, $CO$ and $N_2$ are shown reaching peak values near 160\,km, i.e., the periapsis altitude of MAVEN. In view of the interesting result on the day-to-day variation of thermosphere/exosphere of $Ar$ densities as shown in figure-3, a similar analysis is carried out for other constituents by using only the MAVEN data having more observation points with 4-5 daytime passes per day during June 1-15, 2018. These density profiles of the major gas constituents on different days of June 2018 are shown in the figure-4. It can be seen that all the gas concentrations show day-to-day variations which is very striking for $CO_2$ and $O$. A gradual rise of the $CO_2$-$O$ cross-over altitude from $\sim$200-230\,km between June 3-13, 2018 is quite clear from the figure. The figure also shows the scale for the range of $\sim$10$^{^\circ}$ variation in SZA during these measurements.

\begin{figure}
\begin{center}
	{\includegraphics[width=0.9 \textwidth]{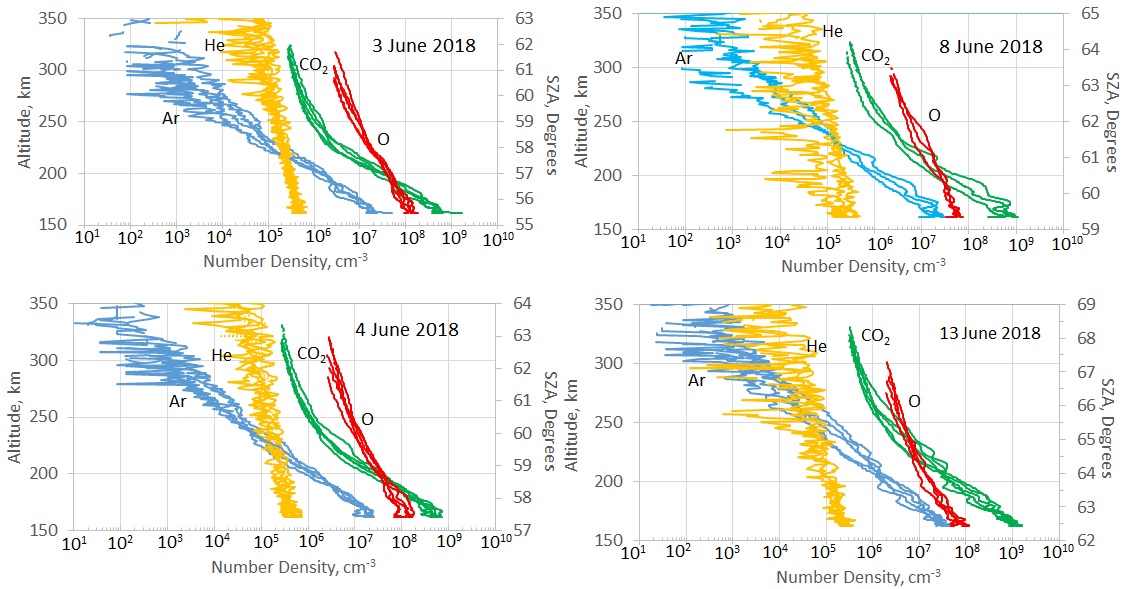}}
	\caption{Density profiles of the gas constituents on a few selected days of June 2018 corresponding to, on an average five daytime orbital passes through the periapsis altitude, the measured altitude profiles have been included to indicate the daily spread of the density profiles apart from showing its day-to-day variations.}	
	\label{fig004}
\end{center}
\end{figure}

Focussing on $CO_2$ and $O$ variation we plot only these two profiles in figure-5. The observed clear reversal between the increase of $CO_2$ density and the decrease of $O$ density for the same period can be noted. To statistically verify this anti-correlation between $CO_2$ and $O$ densities, the standard errors from standard deviations of mean values at 95\% confidence level are computed and the average error profiles are also shown in figure-5. It is noticed that mean standard errors are at least one order of magnitude lower than the range of the observed day-to-day variation. A detailed linear regression analysis has been carried out using the solar energetic H-ion bulk velocities and the solar EUV radiation fluxes as the independent variables and the densities of $CO_2$ and $O$ as dependent variables to determine various correlation coefficients and their statistical significance. The results of this analysis summarised in Table-1 are discussed after presenting the observed variations of the solar EUV and solar wind H-ion velocities/fluxes.

\begin{figure}
	\begin{center}
	{\includegraphics[width=0.75 \textwidth]{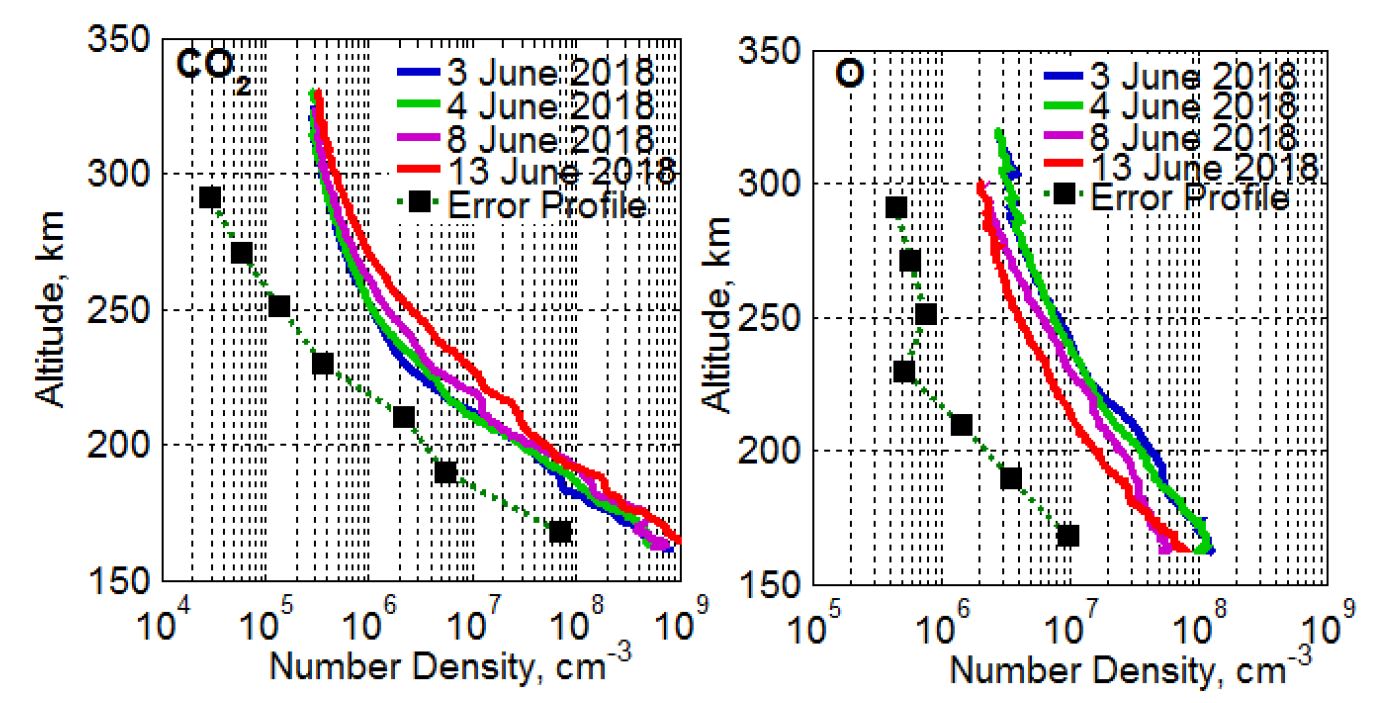}}	
	\caption{Daytime mean density profiles of $CO_2$ and $O$ on four selected days, i.e., 3, 4, 8 and 13 June 2018, using NGIMS observations are shown. The computed mean standard error curves at 95\% confidence level for the two constituents are also plotted. Same colour coding representing different days of observations has been used in both $CO_2$ and $O$ profiles for highlighting the opposite trends between the two constituents. An animated video showing the detailed dynamics of the  day-to-day density profiles of $CO_2$ and $O$ for 14  consecutive days (taking the densities derived from the first morning orbit on each day) in sequence is submitted for online viewing. There are 3 more such videos (not included here) taking the 2nd, 3rd and the 4th orbits from each of the 14 days of data. The results showed similar opposite trends of $CO_2$ and $O$ densities.}
\label{fig005}
	\end{center}
\end{figure}

It is known that solar EUV radiation is absorbed in the thermosphere of Mars that heats it up while producing photochemical and ion-chemical products [26]. The absorbed solar EUV flux would increase thermospheric photodissociation of abundant $CO_2$  resulting in the relative increase of $O$ densities and decrease in $CO_2$ densities. Many authors have developed 1D photochemical models which provide accurate estimates of the distribution of neutral and ion species between 100-300\,km of Martian upper atmosphere [27-29]. 

For a quick appraisal of our results, we have considered main contending reactions as: 

\noindent (i) ${CO_2 + h\nu \rightarrow CO + O (^3P)\; (\lambda < 2274  \dot{A}}$),  (ii) ${CO + O + M \rightarrow CO_2 + M}$ where M is the usual third body to catalyse the reaction. The recombination reaction (ii) shown above being very slow in upper thermosphere and exosphere, $O$ densities build up at the cost of $CO_2$ above $\sim$200\,km. The high energy EUV photons can also ionise $CO_2$, $O$ and $Ar$ with $\lambda$ < 902, 911 and 783 $\dot{A}$ respectively. The equilibrium condition densities of these ionic species may vary with EUV fluxes with changing solar activity affecting the neutral densities. An upgraded 1D chemical model would be required to quantify the solar activity caused variation of densities of both neutral and ionic species. By taking two more reactions: (iii) ${O + O + M \rightarrow O_2 + M}$ and (iv) ${O + O_2 + M \rightarrow O_3 + M}$, and also neglecting photoionisation reactions and using sample values of initial densities, reaction rate coefficients and the EUV fluxes we computed the density changes of $CO_2$ and $O$ for a 30\% increase in solar EUV fluxes. The result showed that for a 30\% increase in EUV flux there is a 22\% decrease in $CO_2$ density and a 29\% increase in $O$ density around 200\,km altitude. While the calculated densities do support the observed anticorrelation between $CO_2$ and $O$ densities into the exosphere, a full 1D model would provide more accurate estimates. More details on the computations using our simplified model are provided in Appendix-1: Details of the Computation of the Effect of EUV on $CO_2$ Dissociation. As a full 1D model with all possible reactions is under preparation by our group, the details in the Appendix-1 explain the basic elements of a simplified 1D model which is used to derive the result on the effect of increased EUV flux on $CO_2$ dissociation rates (please see, for example, Fox [14]; Yung and DeMore [30]).}

To find out the real situation of EUV fluxes, the EUVM experiment data derived from MAVEN mission, is presented as a contour plot of the integrated solar EUV flux ($\lambda$=17-22\,nm) during the period June 1-15, 2018 with days on the x-axis and hours of each day on the y-axis, and given in figure-6. It can be seen that there are periods of slightly higher solar activity (in terms of solar EUV flux increases) during the first week as compared to the second week of June 2018. The maximum variation of this EUV flux is only about 10\%. From the above mentioned model computations, it is clear that such EUV flux changes would produce less than 10\% in the $CO_2$/$O$ densities.

Mars does not have a global, well-structured and strong magnetosphere. But the sub-solar Induced Magnetosphere Boundary (IMB) of Mars is typically located at an altitude of 650\,km. Additional kinetic pressure of ions may push the IMB closer [31] to enable enhanced  Mars-solar wind interaction. So, during higher solar activity the $H^+$ ions could penetrate deeper and interact with the exospheric constituents leading to dissociation of $CO_2$ causing increasing $O$ densities and decreasing $CO_2$ densities.

The solar wind energetic protons ($H^+$ ion) can directly interact with thermosphere-exosphere neutral gas constituents to effect dissociation of $CO_2$ in a similar way as the solar EUV radiation [32] except in case of the former, the particle kinetic energy instead of the energy of photons would determine the magnitude of ion impact dissociation of $CO_2$. Dong et al. [33] have studied the roles of the thermosphere and exosphere on the Martian ionospheric structure and ion escape rates in the process of the solar wind-Mars interaction. The cold thermospheric oxygen atom, however, is demonstrated to be the primary neutral source for $O^+$ ion escape in the thermosphere during the relatively weak solar cycle 24.  Here we show the variation of the $H^+$ ion bulk flow velocities measured near MAVEN spacecraft using SWIA payload onboard MAVEN spacecraft (figure-7). The contours show that the enhanced solar wind velocities are observed in the first week of the chosen period compared to its second week with a slow downward gradient towards the middle of June 2018. But in this case the enhancement is about a factor of 4 or more as compared to the 10\% increase in EUV fluxes. This four-fold increase in proton velocities and an order of magnitude enhancement in $H^+$ particle fluxes (not shown here) are likely to contribute to additional dissociation of $CO_2$ leading to enhanced $O$ during the beginning of the first week of June 2018 and its tapering off later. Krasnopolsky [28] produced a self-consistent model for 11 neutral and 18 ion species from 80 to 300\,km on Mars by solving the continuity equations including ambipolar diffusion for ions. We have carried out calculation using the mechanism of hydrogen ion collisional dissociation of $CO_2$ as follows: ${CO_2 + H^{+*}\rightarrow CO + O + H^+}$ where $H^{+*}$ is the energetic $H^+$ or proton, and the excess kinetic energy is used up in dissociating $CO_2$. The $H^+$ in the product side has less energy and is no longer denoted with the asterisk. We also use the theory of chemical equilibrium taking into account the key recombination equations mentioned earlier. The relative change in densities of $CO_2$ and $O$ for a fixed $H^+$ flux of $10^8$ $m^{-2}$ $s^{-1}$ and increasing bulk velocity from 100 to 400\,km/s have been calculated by incorporating the reaction rate coefficients, collisional cross-sections etc. The details of this model computations are not given here for the sake of brevity. However we have included in Appendix-2, a method of computing solar wind $H^{+*}$ impact dissociation of $CO_2$ using our simple model (as  examples please see, Johnson [34] and Luhmann et al. [35]) for insight into basic computations/model). The result showed a decrease in $CO_2$ density and an increase in $O$ density. For a sample initial $CO_2$ and $O$ density, the percentage decrease in $CO_2$ density and percentage increase in $O$ density were found to be 72\% and 106\%, respectively. We also find that by including the effect of an order of magnitude change in the flux value of solar wind $H^{+*}$ particles $CO_2$ will be completely dissociated reducing it to zero density, and $O$ density will be increased by 1060\%. So, it is possible to get an increase of $O$ density by a factor of two or more at the cost of the depletion of $CO_2$ by similar magnitude through the collisional dissociation by the solar wind $H^+$ ion. However, the details of distribution of enhanced particle flux within the velocity range of 100-400\,km/s need to be  incorporated for more accurate results. This simple model would need to be upgraded to consider many other possible reactions of production and loss of $CO_2$ and $O$. The role of suprathermal atoms and molecules due to the control of induced electric fields and related phenomena affecting the slow escape of lighter neutral species such as $H$, $O$ are not considered here as we dealt with the equilibrium condition densities of different gas constituents. 

This is one likely cause which can explain the observed increasing/decreasing densities of $CO_2$/$O$ as the days progressed from 1st June to 15th June, 2018. It is important that further evidence of this phenomenon of dominance of solar charged particle radiation in controlling the day-to-day densities of the major gaseous concentration profiles in the thermosphere/exosphere would need to be provided. However presently there is a lack of simultaneous data from MOM and MAVEN. Further confirmation of this preliminary result may be obtained when adequate data becomes available from both the Mars orbiters and upcoming Mars missions.

\begin{figure}
\begin{center}
	\includegraphics[width=0.75 \textwidth]{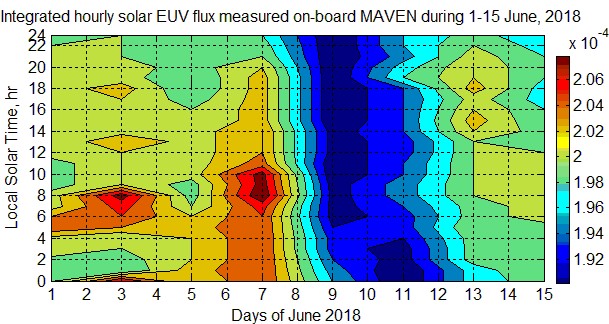}
	\caption{Variation of hourly solar EUV integrated fluxes ($\lambda$ = 17-22\,nm) during  June 1-15, 2018 using EUVM data downloaded from the MAVEN Science Data Centre. Some of the missing data points have been filled by using linear interpolation to ensure correct contour values. The EUV flux has varied between 19$\times$10$^{-5}$ to 21$\times$10$^{-5}$ W\,m$^{-2}$ or by only $\approx$10\% of the mean value.}
\label{fig006}
\end{center}
\end{figure}

\begin{figure}
\begin{center}
	{\includegraphics[width=0.75 \textwidth]{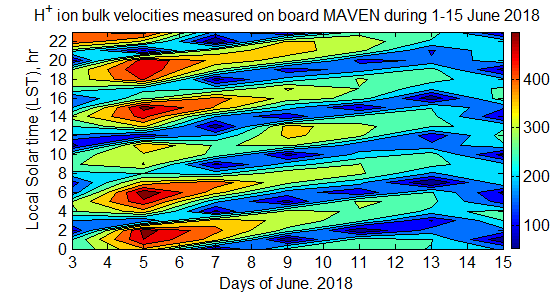}}
	\caption{Solar wind driven $H$-ion bulk velocities measured at MAVEN spacecraft around Mars by SWIA payload. The hourly means of the velocity values have been computed from every 8\,s data points and plotted for alternate days between June 3-15, 2018. While SWIA measures the velocities up to 1000 km\,s$^{-1}$, the magnitude of velocities ranges between $\sim$100-500 km~s$^{-1}$, as shown through the colour codes. The contours of $H$-ion velocities complement the electron/$H$-ion energies and densities measured using the SEP instrument onboard MAVEN.}	
\label{fig007}
\end{center}
\end{figure}
\begin{table}
	\centering
	\caption{Parameters derived from linear regression analysis of dependent variables $CO_2$ and $O$ densities ($y_1$ and $y_2$) and independent variables solar wind $H^+$ ion velocities and solar EUV radiation flux ($x_1$ and $x_2$)}
    
	\label{Table:1}
	\begin{tabular}{lllll}\\
    
		\hline
		Regression Parameter & $y_1$ \& $x_1$ & $y_2$ \& $x_1$ & $y_1$ \& $x_2$ & $x_1$ \& $x_2$ \\
		\hline
		R & -0.80 & 0.55 & -0.43 & 0.41 \\	
		R Square & 0.64 & 0.30 & 0.19 & 0.17\\
		Standard error  & 3.3 & 0.8 & 5.0 & 0.9 \\
		\hline
	\end{tabular}\\
    Note: Standard errors are given for the lowest height in Number Densities ($\times 10^7$)/cc
\end{table}	

The results of the linear regression analysis are summarised in Table-1. Three relevant parameters i.e., the signed (positive or negative) value of the correlation coefficients (R), fraction of a total number of observation points falling on the linear regression line (R-squared), and the standard errors in densities at 95\% confidence level are given in units of number of atoms/molecules per cc as averaged for the periapsis altitude. It can be seen from the table that the most significant high R (correlation coefficient) value of -0.8 (negative correlation) with 64\% matching points (R-squared) falling on the regression line and lower than one order of magnitude standard error in densities is between the variation of $CO_2$ densities with the solar wind $H$-ion velocities during the observation period. The high negative correlation means that the $CO_2$ densities for the height range of 160-300\,km decreased with the increasing solar wind $H$-ion velocities. The same parameters between $O$ densities and $H$-ion velocities are 0.55 (positive correlation) with 30\% matching points and less than one order of magnitude standard error. The positive correlation here indicates that the $O$ densities increased with the increase of the solar wind $H$-ion velocities. The same regression parameters show somewhat similar trends with solar EUV radiation i.e., decreasing/increasing $CO_2$ and $O$ densities with increasing solar EUV radiation but with poor correlation coefficients (-0.43 \& 0.40, respectively) and relatively low statistical significance.

\section{Conclusion}
Analysis of near-simultaneous mass spectrometric observations by the two orbiters MOM and MAVEN showed an  increasing trend of Argon ($Ar$) density profiles in the thermosphere-exosphere ($\sim$150-300\,km) region during the period June 1-15, 2018 when the solar EUV radiation flux did not show any appreciable difference. Using the daily mean data from every five hourly observations by NGIMS/MAVEN on exospheric neutral densities and composition, the same method of analysis is extended to other primary atmospheric constituents (e.g., $CO_2$, $O$,  etc.) during the same period. An important finding is that the densities of $CO_2$ profiles progressively increased and that of $O$ decreased during 1 June to 15 June, 2018.

Daily Data from other MAVEN instruments such as EUVM and SWIA, were used to check the variation of mean solar EUV radiation fluxes at 17-22\,nm wavelength band and the hourly average values of the solar wind $H$-ion bulk velocities respectively. It is noted that while the EUV radiation decreased only by $\sim$10\%, the solar wind ion velocities decreased by four orders of magnitude during June 1-15, 2018. This indicates that the increase in solar wind energetic particle velocities in the first week of June would have decreased $CO_2$ densities through charged particle impact dissociation producing more $O$ densities thus showing an anti-correlation between $CO_2$ and $O$ densities. A linear regression analysis and statistical significance test support this conclusion of the possible cause of the observed day-to-day variation and the anti-correlation between $CO_2$ and $O$ densities in the upper thermosphere and lower exosphere of Mars. Further, a very simplified 1D model run incorporating the steady state chemical equilibrium condition between the production by the solar wind charged particle impact dissociation of $CO_2$ to produce $O$ and a few relevant recombination/loss processes provided sample results which support the observations. For an enhanced velocity of solar wind $H^+$ ion from 100 to 400\,km/s the density decrease and increase for $CO_2$ and $O$ respectively show variation by a factor of $\sim$2, which can be even more if we consider the concurrent differential enhancement of $H^+$ flux values. 

These preliminary results would need further confirmation with additional and preferably simultaneous observations by different Mars Orbiters and verification by a combined and improved 1D photochemical and ion-chemical model incorporating a built-in routine for computing the dissociation and ionisation by the impact of solar wind charged particle radiation. 

\section*{Acknowledgments}
The authors acknowledge the support from team members of the MENCA-MOM instrument group at Space Physics Laboratory, ISRO for providing the calibration data. This research work was funded by the Indian Space Research Organisation (ISRO) through the MOM-AO project on Observation and Modelling Studies of the Atmospheric Composition of Mars (OMAC), vide reference number: ISRO:SPL:01.01.33/16/2016.

\section*{Data Availability}
The NGIMS, SWIA, EUVM and SEP datasets of MAVEN used for this study are publicly available on MAVEN Science Data Center at the Laboratory for Atmospheric and Space Physics (LASP)-CU Boulder (\url{https://lasp.colorado.edu/maven/sdc/public/}) as well as at the Planetary Data System of NASA (\url{https://pds.nasa.gov}). The MAVEN mission is supported by NASA through the Mars Exploration Program. The MENCA data from the MOM mission has been archived at the Indian Space Science Data Center, Bengaluru (\url{https://www.issdc.gov.in/}).\\

\section*{Author Contribution} 
The research was carried out jointly with shared responsibilities, and both authors contributed to the manuscript. Kamsali Nagaraja and S.C. Chakravarty worked together to develop the concept, methodology, and formal analysis and conducted the investigation. Kamsali Nagaraja and S.C. Chakravarty were also involved in the initial drafting, revision, and editing of the manuscript in its entirety. Kamsali Nagaraja served as the Principal Investigator for an ISRO project. S.C. Chakravarty handled the resources and supervised the work.

\section*{References}
\begin{itemize}
	\item[1)] Martinez GM, Newman CN, De Vicente-Retortillo A, Fischer E, Renno NO, Richardson MI, Fairén AG, Genzer M, Guzewich SD, Haberle RM, Harri AM. The modern near-surface Martian climate: a review of in-situ meteorological data from Viking to Curiosity. Space Science Reviews. 2017; 212: 295-338.
	\item[2)] Nier AO, McElroy MB. Structure of the neutral upper atmosphere of Mars: Results from Viking 1 and Viking 2. Science. 1976; 194(4271): 1298-300.
	\item[3)] Nier AO, Hanson WB, Seiff A, McElroy MB, Spencer NW, Duckett RJ, Knight TC, Cook WS. Composition and structure of the Martian atmosphere: Preliminary results from Viking 1. Science. 1976; 193(4255): 786-8.
	\item[4)] Owen T, Biemann K, Rushneck DR, Biller JE,  Howarth DW, Lafleur AL. The composition of the atmosphere at the surface of Mars.  Journal of Geophysical Research. 1977; 82: 4635-39.
	\item[5)] Jakosky BM, Lin RP, Grebowsky JM, Luhmann JG, Mitchell DF, Beutelschies G, Priser T, Acuna M, Andersson L, Baird D, Baker D. The Mars atmosphere and volatile evolution (MAVEN) mission. Space Science Reviews. 2015; 195: 3-48.
	\item[6)] Arunan S, Satish R. Mars Orbiter Mission spacecraft and its challenges. Current Science. 2015; 109: 1061-69.
	\item[7)] Valeille A, Combi MR, Bougher SW, Tenishev V, Nagy AF. Three‐dimensional study of Mars upper thermosphere/ionosphere and hot oxygen corona: 2. Solar cycle, seasonal variations, and evolution over history. Journal of Geophysical Research: Planets. 2009; 114(E11).
	\item[8)] Medvedev AS, González‐Galindo F, Yiğit E, Feofilov AG, Forget F, Hartogh P. Cooling of the Martian thermosphere by CO2 radiation and gravity waves: An intercomparison study with two general circulation models. Journal of Geophysical Research: Planets. 2015; 120(5): 913-27.
	\item[9)] Halekas JS, Ruhunusiri S, Harada Y, Collinson G, Mitchell DL, Mazelle C, McFadden JP, Connerney JE, Espley JR, Eparvier F, Luhmann JG. Structure, dynamics, and seasonal variability of the Mars‐solar wind interaction: MAVEN Solar Wind Ion Analyzer in‐flight performance and science results. Journal of Geophysical Research: Space Physics. 2017; 122(1): 547-78.
	\item[10)] Bougher SW, Pawlowski D, Bell JM, Nelli S, McDunn T, Murphy JR, Chizek M, Ridley A. Mars Global Ionosphere‐Thermosphere Model: Solar cycle, seasonal, and diurnal variations of the Mars upper atmosphere. Journal of Geophysical Research: Planets. 2015a; 120(2): 311-42.
	\item[11)] Mahaffy PR, Benna M, Elrod M, Yelle RV, Bougher SW, Stone SW, Jakosky BM. Structure and composition of the neutral upper atmosphere of Mars from the MAVEN NGIMS investigation. Geophysical research letters. 2015a;  42(21): 8951-7.
	\item[12)] Nagaraja K, Basuvaraj PK, Chakravarty SC, Kumar P. Study of exospheric neutral composition of mars observed from Indian mars orbiter mission. New Astronomy. 2020; 77: 101349. \\ doi.org/10.1016/j.newast.2019.101349
	\item[13)] Sarris TE. Understanding the ionosphere thermosphere response to solar and magnetospheric drivers: status, challenges and open issues. Philosophical Transactions of the Royal Society A. 2019; 377(2148): 20180101. https://doi.org/10.1098/rsta.2018.0101
	\item[14)] Fox JL. The production and escape of nitrogen atoms on Mars. Journal of Geophysical Research: Planets. 1993; 98(E2): 3297-310, doi:10.1029/92JE02289.
	\item[15)] Venkateswara Rao N, Gupta N, Kadhane UR. Enhanced densities in the Martian thermosphere associated with the 2018 planet‐encircling dust event: Results from MENCA/MOM and NGIMS/MAVEN. Journal of Geophysical Research: Planets. 2020; 125(10):e2020JE006430. \\ https://doi.org/10.1029/2020JE00643
	\item[16)] Bhardwaj A, Mohankumar SV, Das TP, Pradeepkumar P, Sreelatha P, Sundar B, Nandi A, Vajja DP, Dhanya MB, Naik N, Supriya G. MENCA experiment aboard India's mars orbiter mission. Current Science. 2015; 109: 1106-13. 
	\item[17)] Bhardwaj A, Thampi SV, Das TP, Dhanya MB, Naik N, Vajja DP, Pradeepkumar P, Sreelatha P, Supriya G, K AJ, Mohankumar SV. On the evening time exosphere of Mars: Result from MENCA aboard Mars Orbiter Mission. Geophysical Research Letters. 2016; 43(5): 1862-7.
	\item[18)] Bhardwaj A, Thampi SV, Das TP, Dhanya MB, Naik N, Vajja DP, Pradeepkumar P, Sreelatha P, K AJ, Thampi RS, Yadav VK. Observation of suprathermal argon in the exosphere of Mars. Geophysical Research Letters. 2017; 44(5): 2088-95.
	\item[19)] Mahaffy PR, Benna M, King T, Harpold DN, Arvey R, Barciniak M, Bendt M, Carrigan D, Errigo T, Holmes V, Johnson CS. The neutral gas and ion mass spectrometer on the Mars atmosphere and volatile evolution mission. Space Science Reviews. 2015b; 195: 49-73.
	\item[20)] Bougher S, Jakosky B, Halekas J, Grebowsky J, Luhmann J, Mahaffy P, Connerney J, Eparvier F, Ergun R, Larson D, Mcfadden J. Early MAVEN Deep Dip campaign reveals thermosphere and ionosphere variability. Science. 2015b; 350(6261): aad0459.
	\item[21)] Eparvier FG, Chamberlin PC, Woods TN, Thiemann EM. The solar extreme ultraviolet monitor for MAVEN. Space Science Reviews. 2015; 195: 293-301.
	\item[22)] Larson DE, Lillis RJ, Lee CO, Dunn PA, Hatch K, Robinson M, Glaser D, Chen J, Curtis D, Tiu C, Lin RP. The MAVEN solar energetic particle investigation. Space Science Reviews. 2015; 195: 153-72.
	\item[23)] Halekas JS, Taylor ER, Dalton G, Johnson G, Curtis DW, McFadden JP, Mitchell DL, Lin RP, Jakosky BM. The solar wind ion analyzer for MAVEN. Space Science Reviews. 2015; 195: 125-51.
	\item[24)] Jain SK, Stewart AIF, Schneider NM, Deighan J, Stiepen A, Evans JS, Stevens MH,  Chaffin MS, Crismani M, McClintock WE,  Clarke JT,  Holsclaw GM,  Lo DY,  Lefèvre F,  Montmessin F,  Thiemann EMB,  Eparvier F,  Jakosky BM. The structure and variability of Mars upper atmosphere as seen in MAVEN/IUVS dayglow observations. Geophysical Research Letters. 2015; 42: 9023–30, doi:10.1002/2015GL065419.
	\item[25)] Kuroda T, Medvedev AS, Yiğit E, Hartogh P. Global distribution of gravity wave sources and fields in the Martian atmosphere during equinox and solstice inferred from a high-resolution general circulation model. Journal of the Atmospheric Sciences. 2016; 73(12): 4895-909. doi:10.1175/jas-d-16-0142.1
	\item[26)] Fox JL, Hac AB. The escape of O from Mars: Sensitivity to the elastic cross sections. Icarus. 2014; 228: 375-85.
	\item[27)] McElroy MB, Kong TY, Yung YL. Photochemistry and evolution of Mars' atmosphere: A Viking perspective. Journal of Geophysical Research. 1977; 82(28): 4379-88, doi:10.1029/js082i028p04379
	\item[28)] Krasnopolsky VA. Mars' upper atmosphere and ionosphere at low, medium, and high solar activities: Implications for evolution of water. Journal of Geophysical Research: Planets. 2002; 107(E12): 5128,  doi:10.1029/2001JE001809
	\item[29)] González‐Galindo F, López‐Valverde MA, Angelats i Coll M, Forget F. Extension of a Martian general circulation model to thermospheric altitudes: UV heating and photochemical models. Journal of Geophysical Research: Planets. 2005; 110(E9): E09008, doi:10.1029/2004JE002312
	\item[30)] Yung YL, DeMore WB. Photochemistry of Planetary Atmospheres. 1999: Oxford University Press. ISBN: 9780195105018, 1-480.
	\item[31)] Edberg NJ, Auster U, Barabash S, Bößwetter A, Brain DA, Burch JL, Carr CM, Cowley SW, Cupido E, Duru F, Eriksson AI. Rosetta and Mars Express observations of the influence of high solar wind pressure on the Martian plasma environment. InAnnales Geophysicae. 2009; 27(12): 4533-45.
	\item[32)] Zurek RW, Tolson RA, Bougher SW, Lugo RA, Baird DT, Bell JM, Jakosky BM. Mars thermosphere as seen in MAVEN accelerometer data. Journal of Geophysical Research: Space Physics. 2017; 122(3): 3798-814, doi:10.1002/2016JA023641	
	\item[33)] Dong C, Bougher SW, Ma Y, Lee Y, Toth G, Nagy AF, Fang X, Luhmann J, Liemohn MW, Halekas JS, Tenishev V. Solar wind interaction with the Martian upper atmosphere: Roles of the cold thermosphere and hot oxygen corona. Journal of Geophysical Research: Space Physics. 2018; 123(8): 6639-54, doi:10.1029/2018JA025543
	\item[34)] Johnson RE. Energetic Charged-Particle Interactions with Atmospheres and Surfaces. Berlin; Springer-Verlag, ISBN: 0387519084, 1990.
	\item[35)] Luhmann JG, Johnson RE, Zhang MHG.  Evolutionary impact of sputtering of the Martian atmosphere by $O^+$ pickup ions. Geophysical Research Letters. 1992: 19(21), 2151-54, doi:10.1029/92GL02485.	
\end{itemize}

\newpage 
	\section*{Appendix-1: Details of the computation of the effect of EUV on $CO_2$ dissociation}
	
	In our simplified model of EUV-produced dissociation of $CO_2$ in the Martian atmosphere, we consider the following primary photochemical reactions:
	\begin{equation}
	CO_2 + h\nu \rightarrow CO + O \hspace{1cm}(\lambda < 242 \, nm)
	\end{equation}
	
	\begin{equation}
	O + O + M \rightarrow O_2 + M
	\end{equation}
	
	\begin{equation}
	O + O_2 + M \rightarrow O_3 + M
	\end{equation}
	
	\begin{equation}
	CO + O + M \rightarrow CO_2 + M
	\end{equation}
	
	These reactions represent the fundamental processes occurring under the influence of solar EUV radiation. To quantify the impact of an increase in solar EUV flux, we use the basic photochemical rate equation:
	
	\begin{equation}
	rate = k \cdot [A] \cdot J
	\end{equation}
	
	where:
	\begin{itemize}
		\item $rate$ = the rate of the photochemical reaction
		\item $k$ = the rate constant of the reaction
		\item $[A]$ = the concentration of the reactant (in this case, $CO_2$)
		\item $J$ = the photolysis rate, proportional to the solar EUV flux
	\end{itemize}
	
	The density variation of the Martian atmosphere due to dissociation of $CO_2$ by solar EUV radiation can be expressed as:
	
	\begin{equation}
	\rho = (\rho_0) e^{({-n}{\sigma}{F}{t})}
	\end{equation}
	
	where:
	\begin{itemize}
		\item $\rho$ = the density of the Martian atmosphere after dissociation $(kg/m^3)$ 
		\item $\rho_0$ = the initial density of the Martian atmosphere before dissociation $(kg/m^3)$
		\item $n$ = the number density of $CO_2$ molecules in the Martian atmosphere $(m^{-3})$
		\item $\sigma$ = the cross section for dissociation of $CO_2$ by EUV radiation $(m^{2})$
		\item $F$ = the solar EUV flux at Mars $(W/m^{2})$
		\item $t$ = the time elapsed since dissociation $(s)$
	\end{itemize}
	
	For a 30\% increase in the solar EUV flux, the new flux $F'$ can be expressed as:
	
	\begin{equation}
	F^\prime = 1.3 \cdot F
	\end{equation}
	
	Using this increased flux, the revised densities of $CO_2$ and $O$ can be computed by plugging $F^\prime$ into the above equations. Due to space constraints, the detailed numerical computations are omitted here. However, the reader can verify the results by applying the formulas provided (for examples of such models, please see Fox  (1993); Yung  and DeMore (1999)).

	\section*{Appendix-2: Details of the computation of  $H^+$ ion impact dissociation of $CO_2$}
	
	In our simplified model of $H^+$ ion impact dissociation of $CO_2$ in the Martian atmosphere, we consider the following primary reactions:
	
	\begin{equation}
	CO_2 + H^{+*} \rightarrow CO + O + H^+
	\end{equation}
	
	The rate of dissociation ($R_d$) of $CO_2$ by proton impact is given by:
	
	\begin{equation}
	R_d = \sigma \cdot n_{CO_2} \cdot \Phi
	\end{equation}
	
	where:
	\begin{itemize}
		\item $\sigma$ = the cross-section for dissociation $(m^2)$
		\item $n_{CO_2}$ = the number density of $CO_2$ molecules $(m^{-3})$
		\item $\Phi$ = the proton flux $(m^{-2}s^{-1})$
	\end{itemize}
	
	Here, $\Phi$ is related to the bulk velocity ($v$) of the protons by:
	
	\begin{equation}
	\Phi = n_{H^+} \cdot v
	\end{equation}
	
	where:
	\begin{itemize}
		\item $n_{H^+}$ = the number density of protons $(m^{-3})$
		\item $v$ = the bulk velocity of the protons $(m/s)$
	\end{itemize}
	
	For the recombination reaction, we consider:
	
	\begin{equation}
	CO + O + M \rightarrow CO_2 + M
	\end{equation}
	
	The rate of recombination $(R_r)$ is given by:
	
	\begin{equation}
	R_r = k \cdot [CO] \cdot [O] \cdot [M]
	\end{equation}
	
	where:
	\begin{itemize}
		\item $k$ = the recombination rate constant
		\item $[CO]$, $[O]$, $[M]$ = concentrations of CO, O, and a third body M, respectively
	\end{itemize}
	
	The equilibrium densities are determined by balancing the dissociation and recombination rates:
	
	\begin{equation}
	R_d = R_r
	\end{equation}
	
	\textbf{Initial Conditions:}
	\begin{itemize}
		\item Proton flux $\Phi) = (10^8\, m^{-2}s^{-1}$
		\item Bulk velocity $(v) = 100\,km/s$
		\item Initial density of $CO_2$, $n_{CO_2} =n_0$
		\item Initial density of $O$, $n_O = n_1$
	\end{itemize}
	
	\textbf{Dissociation Rate with Increased Flux and Velocity:}
	\begin{equation}
	R_d' = \sigma \cdot n_{CO_2} \cdot \Phi'
	\end{equation}
	
	Given $\Phi' = 10^9$ $m^{-2}s^{-1}$ and $v'=400\,km/s$ \\
	
	\textbf{Revised Densities:}
	\begin{equation}
	New[n_{CO_2}] = n_0 \cdot \exp(-k_d \cdot t)
	\end{equation}
	
	\begin{equation}
	New[n_O] = n_1 + n_0 \cdot (1 - \exp(-k_d \cdot t))
	\end{equation} \\
	
	For an initial proton flux $\Phi = 10^8\,m^{-2}s^{-1}$: $CO_2$ density reduction = 72\%; $O$ density increase = 106\%. \\
	
	For an order of magnitude increase in flux $(\Phi' = 10^9 \,m^{-2}s^{-1} )$: the $CO_2$ density reduction = 100\%; and the $O$ density increase = 1060\%. \\
	
	Thus, the increase in proton flux by an order of magnitude results in the near-complete depletion of $CO_2$ and a significant increase in $O$ density (for examples of such models please see  Johnson (1990);  Luhmann et al. (1992)).

\end{document}